\pdfoutput=1

\documentclass[12pt]{article}\usepackage[]{graphicx}\usepackage[]{color}
\makeatletter
\def\maxwidth{ %
  \ifdim\Gin@nat@width>\linewidth
    \linewidth
  \else
    \Gin@nat@width
  \fi
}
\makeatother

\definecolor{fgcolor}{rgb}{0.345, 0.345, 0.345}

\usepackage{framed}
\makeatletter
 {\par\unskip\endMakeFramed%
 \at@end@of@kframe}
\makeatother

\definecolor{shadecolor}{rgb}{.97, .97, .97}
\definecolor{messagecolor}{rgb}{0, 0, 0}
\definecolor{warningcolor}{rgb}{1, 0, 1}
\definecolor{errorcolor}{rgb}{1, 0, 0}
\usepackage{alltt}
  
\usepackage{tcolorbox}
\usepackage[a4paper,margin=0.75in]{geometry}
\usepackage{amsmath} %Never write a paper without using amsmath for its many new commands
\usepackage{amssymb} %Some extra symbols
\usepackage{makeidx,showidx} %If you want to generate an index, automatically
\usepackage{graphicx} %If you want to include postscript graphics
\usepackage{amsmath,amsfonts,amssymb,amsthm,graphicx, graphics,color,url,lscape,natbib,courier}
\usepackage{multicol}
\usepackage[utf8]{inputenc}
\usepackage{authblk}

\newcommand{\disp}{\displaystyle}
  
\title{\textbf{Understanding Sea Ice Melting \\ via Functional Data Analysis}}
\author{Purba Das, Ananya Lahiri, Sourish Das\footnote{Contact at: sourish@cmi.ac.in}}
\affil{Chennai Mathematical Institute, INDIA}
\date{September 15, 2016}
\IfFileExists{upquote.sty}{\usepackage{upquote}}{}
\begin{document}
\maketitle
  %\begin{multicols}{2}
  
\begin{abstract}
In this article, we considered the problem of sea ice cover is melting. Considering the `satellite passive microwave remote sensing data' as functional data, we studied daily observation of sea ice cover of each year as a  smooth continuous function of time. We investigated the mean function for the sea ice area for following decades and computed the corresponding $95\%$ bootstrap confidence interval for the both Arctic and Antarctic Oceans. We found the mean function for the sea ice area dropped statistically significantly in recent decades for the Arctic Ocean. However,  no such statistical evidence was found for the Antarctic ocean. Essentially, the mean function for sea ice area in the Antarctic Ocean is unchanged. Additional evidence of the melting of sea ice area in the Arctic Ocean is provided by three types of phase curve (namely, Area vs. Velocity, Area vs. Acceleration, and Velocity Vs. Acceleration). In the Arctic Ocean, during the summer, the current decades is observing the size of the sea ice area about 30\% less, than what it used to be during the first decade. In this article, we have taken a distribution-free approach for our analysis, except the data generating process, belongs to the Hilbert space.
\end{abstract}


\section{Introduction}

Recently, the functional data analysis (FDA) is attracting a lot of attention in the scientific community (\cite{RamsaySilverman2005}). With the new digital explosion of data, the FDA is a highly useful tool to understand many complex systems. Functional data analysis is broadly used in biostatistics, growth curve data, longitudinal data analysis, meteorology etc. (\cite{FDAreview2016},\cite{RamsayAFDA2002}). In this paper, we provide statistically significant evidence of melting of the sea ice area in the Arctic Ocean over last decade using FDA. We also provide statistical observation of the sea ice area of the Antarctic Ocean. We treat the satellite passive microwave remote sensing daily data for a year as a functional data for both the Arctic and Antarctic oceans. We also presented a functional measure of severity of the change on previous decade.

The measurement of extending and trend are available for Arctic (\cite{claire1999}).  The satellite evidence of the transformation of the Arctic sea ice cover was presented by \cite{Ola1999}. For the Arctic Ocean, sharp change in sea ice cover over time has been reported in \cite{Comiso2002} , \cite{Stroeve2007},\cite{claire2008}. A review of the effect of Arctic sea ice decline on weather and climate can be seen in \cite{Timo2014}. But very few articles have considered Arctic as well as Antarctic sea ice cover for their observation. Recently (\cite{comiso2016}) provides evidence for asymmetry in the rate of change of sea ice cover melting in the Arctic and Antarctic ocean. After a thorough search, we observed no work considered the data as a smooth functional process over time.

We want to make a critical remark, in all the other paper the authors have discussed accelerated declination of sea ice cover but have not provided any confidence intervals for their analysis. But just looking at the mean curves (or the first derivative) without the confidence interval we can not conclude if statistically significant change has taken place.  In this paper, we presented $95\%$ bootstrap confidence interval for analysis. We use the \texttt{FDA} package(\cite{FDA}) in \texttt{R} for the Fourier smoothing of raw data and for further analysis.

In section \ref{sec2}, we present the data source and discuss the motivation for using the FDA in analyzing sea ice cover. In section \ref{sec3}, we smooth the sea ice area raw data. In section \ref{sec4}, we provide some general observations and summary Statistic. Section \ref{sec5} and \ref{sec6} corresponds to two different Statistical evidence of sea ice area melting of Arctic Ocean over time and some view of Antarctic sea ice area over time. In section \ref{sec7}, we provide a measurement of sea ice area melting.

\section{Functional Data \label{sec2}}

\subsection{Data Source and Description}

\par We have used two different data sources. The Arctic and Antarctic daily sea ice area data are available from National Snow $\&$ Ice Data Center(NSIDC), website \url{https://nsidc.org/data/seaice_index/archives.html}. This data is available from Nov 1978 to Dec 2015. As we want to study the yearly effect we considered the data from $1^{st}$ Jan 1979 - $31^{st}$ Dec 2015($37$ years data) for our analysis for both the Arctic and Antarctic. 

\par Another source of the same data for both Arctic and Antarctic are available from NASA, website \url{http://neptune.gsfc.nasa.gov/csb/index.php?section=59}. This data is available from Nov 1978 to Dec 2013. In both source,from $1^{st}$ Jan 1979 to $20^{th}$ Aug 1987 the sea ice area data are available only on alternate days and from 21st Aug 1987 onward we have daily data. The dataset contains the total sea ice area of the Arctic ocean along with the break up of the nine different seas of the Arctic ocean. 

\subsection{Why Functional Data?}
\begin{figure}[h!]
\begin{center}
\caption{Arctic and Antarctic sea ice area(in million square km) over days for 37 years (observed data)}
\label{fig_raw_data}
\includegraphics[height=3.9in]{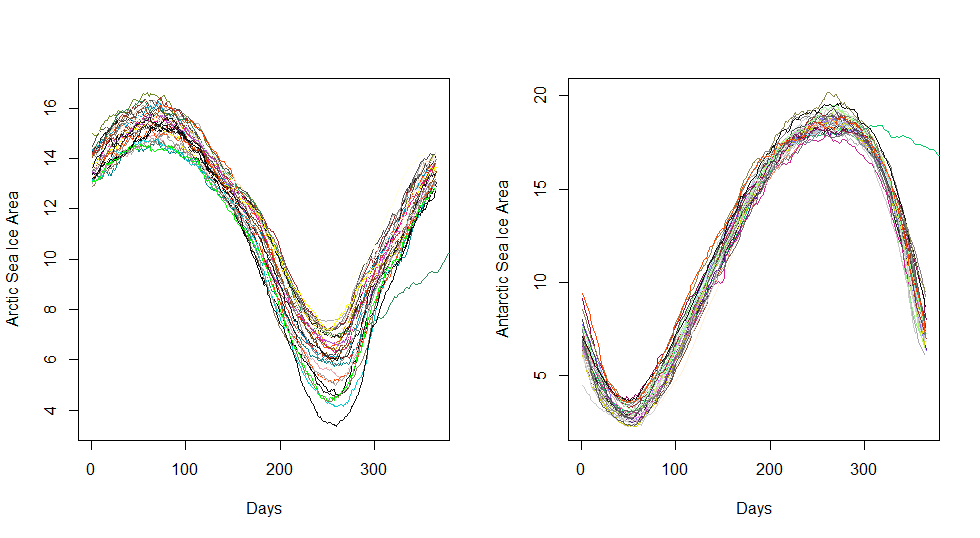}
\end{center}
\end{figure}

The functional data assumes each sample element is a smooth function. The variable sea ice area takes value continuously. For our purpose, the sea ice area over a particular year is considered a continuous and smooth function of time. We plot the $37$ years raw sea ice area functional data in the figure (\ref{fig_raw_data}). In our analysis, we have total $37$ years each corresponds to a functional curve. The functional curve for all the 37 years has the same trend. Any distributional assumption is not made in this work.

\subsection{Fourier Smoothing of Functional Data\label{sec3}}

From the figure (\ref{fig_raw_data}), we see that the observed sea ice area has some local randomness in it. For our analysis, for each year $i \in \{1,2...37\}$, we consider the model as follows:
\begin{equation*}
y_{j} = x(t_{j}) + e_{j}, ~~~ j = 1(1)n_i
\end{equation*}
where the error term $e_{j}$ contributes to the roughness to the observation, $y=(y_{1},y_{2},...y_{n_i})$ is the observed vector with $n_i$ is the number of data point for $i^{th}$ year, $x(t_{j})$ is the functional value of the continious smooth funtion $x(t)= \disp\sum_{k=1}^{p} c_{k}\phi_{k}(t) $ evaluated at the time point $t_{j}$. We choose $x$ as a fourier smooth function. Here $p$ is a fixed constant. As Fourier basis, the number of basis element has to be odd and 
\begin{eqnarray*}
\phi_{k}(t) = \bigg\{
\begin{array}{cc}
 1,& \text{if } \text{k = 1,}\\
\sin(\frac{k}{2} \omega t) ,& \text{if } \text{k is even,}\\
\cos(\frac{k-1}{2} \omega t),              & \text{otherwise.}
\end{array}
\end{eqnarray*}
Now for our analysis, define, $T=(i)_{i=1}^{365}$, a $365$ length vector (so, $n$ is $365$ for our case). The observed data vector $y$ is denoted as $y=(y_{T[i]})_{i=1}^{365}$. The model of $y_{T[i]}$ is 
\begin{equation*}
y_{T[j]} =x(T[j]) + e_{T[j]}
\end{equation*}

%  \\  For Fourier smooth, for each year $i \in \{1,2...37\}$ we want to estimate a continuous differentiable function $x_{i}(t)$ $  \forall$ $ t \in [0,365]$, such that the observed $365$ dimensional data vector $y_{i}=(y_{1},y_{1},...y_{365})^{t} $ for  $i^{th}$ year is $y_{i} =x_{i}(t) + e_{i}$ with $e_{i} = (e_{i1},e_{i2},...e_{i365})$.\\
%  We estimate $x(t)$ as 
%  \begin{center}
%  $x(t)= \disp\sum_{k=1}^{p} c_{k}\phi_{k}(t) $. 
%  \end{center}
%  Where, p is a fixed odd(as Fourier basis, the no of basis element has to be odd) constant and \begin{center}
%  \[
%     $\phi_{k}(t) = $ 
%  \begin{cases}
%      \disp 1,& \text{if } \text{k = 1}\\
%      \sin(\frac{k}{2} \omega t) ,& \text{if } \text{k is even}\\
%      \cos(\frac{k-1}{2} \omega t),              & \text{otherwise}
%  \end{cases}
%  \]
%  \end{center}%
%  Now whenever $p \rightarrow\infty $ the $e_{ij}\rightarrow 0$ $ \forall i,j$ ie. we end up modeling the error also and $p=1$ implies estimate y through a constant function.

\begin{figure}[h!]
\begin{center}
\caption{Arctic and Antarctic sea ice area (in million square km) after smoothing over days for $37$ years}
\label{fig_smooth_data}
\includegraphics[height=3.9in]{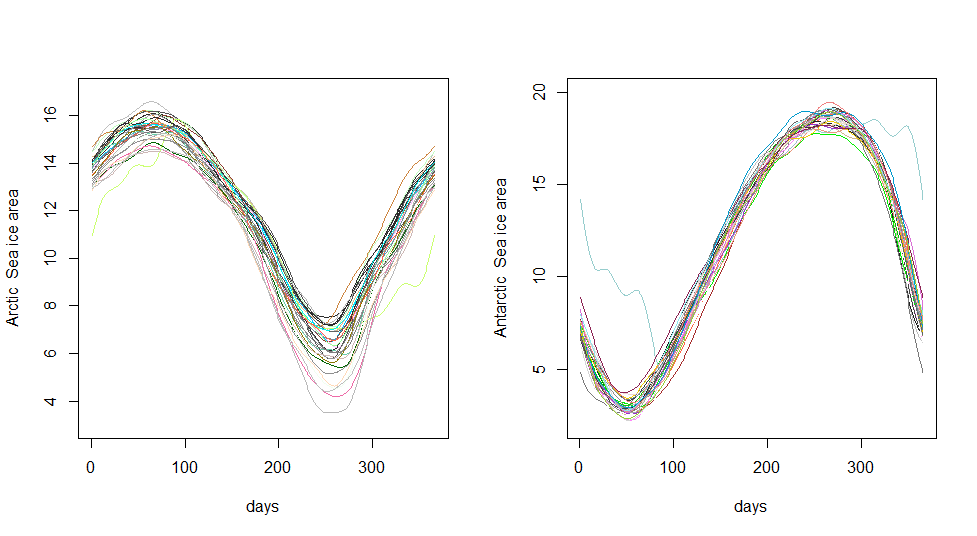}
\end{center}
\end{figure}

In figure \ref{fig_smooth_data} we present the curves of the sea ice area using the Fourier basis smoothing. We compute the coefficients $c_{k}$'s by weighted least square method using \texttt{FDA} package in \texttt{R} \cite{FDA}.

\subsection{Number of Basis Selection}
To find the optimal number of basis we model the mean squared error(MSE), i.e. we calculate\\
$$
MSE_{i}^{p}=\frac{1}{n_i} \sum_{j=1}^{n_i} e_{ij}^2
$$ 
where, $i$ $\in \{1,2,...37\}$  and $p$ (odd integer) is the number of basis we used to estimate $e_{ij}$, $p$ is odd because number of basis elements in Fourier basis can not be even, throughout the paper, we assume the mean and the variance of $e$ exists..

\begin{figure}[h!]
\begin{center}
\caption{$\widehat{MSE}$ Vs. number of fourior basis (Gray line represents no. of basis=$21$)}
\label{fig3}
\includegraphics[height=3in]{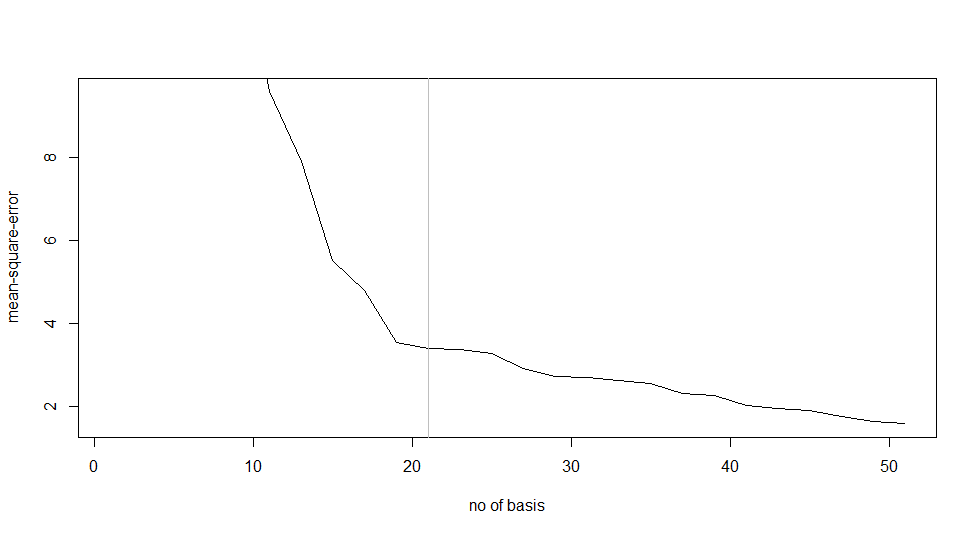}
\end{center}
\end{figure}

\begin{figure}[h!]
\begin{center}
\caption{First difference of $\widehat{MSE}$ (Gray lines corresponds to no. of basis=21 and  $\widehat{MSE}=0$ ) }
\label{fig4}
\includegraphics[height=3in]{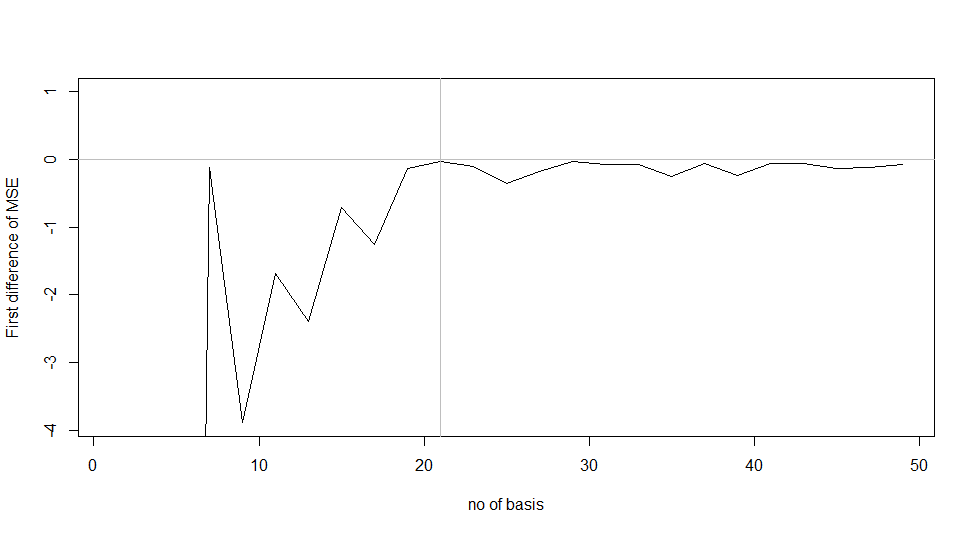}
\end{center}
\end{figure}

Now we compute the $MSE_{i}^{p}$ for each $i$ $\in \{1,2,...37\}$ given a fixed $p$. We define 
\begin{equation*}
\widehat{MSE^{p}} := \frac{1}{37}\sum_{j=1}^{37}MSE_{i}^{p} ,
\end{equation*}
where $p$ is a fixed odd integer. Now we took $p$ to be all odd integers between $1$ and $51$ and find $\widehat{MSE^{p}}$ for each $26$ values of $p$ and plot them in figure (\ref{fig3}). Directly from the plot, we can observe that the slope is changing at $p=21$, but we can not argue that $p=21$ is the optimal choice for number of basis. Hence, we study the plot of first difference of $\widehat{MSE}$ in figure (\ref{fig4}).

Though the graph of $\widehat{MSE}$ (figure \ref{fig3}) is strictly decreasing, but the graph of first difference of $\widehat{MSE}$ plot (figure \ref{fig4}) is more or less constant after $21$. If we go on increasing the number of basis, we observe the estimate overfits. After exceeding the 21 number of basis, we observe that the change is very small and almost close to zero. Hence, from here onwards, in the rest of the paper, we use $21$ as the optimal choice for the number of Fourier basis elements.

\textbf{Remark}: Both the plots in figure  (\ref{fig3},\ref{fig4}) are corresponding to the Arctic sea ice area. For the Antarctic sea ice area also the value of $p$ turns out to be 21. We skip the plot in this paper for space constraint.

\section{General Observation and Summary Statistics\label{sec4}}

For the Arctic Ocean, the sea ice area reaches the maximum between $5^{th}$ to $9^{th}$ March and it reaches minimum $10^{th}$ to $14^{th}$ September. This highest and lowest sea ice area timing are around 30 days later when the temperatures reach minimum and maximum respectively\footnote{Source: mean temperature for the north of the 80 degrees north cane be found here: \url{http://ocean.dmi.dk/arctic/meant80n.uk.php}}. So the time lag between extreme temperature and extreme sea ice area of Arctic Ocean is around 30 days. 

For Antarctic Ocean the sea ice area reaches the maximum between $22^{nd}$ to $26^{th}$ September and it reaches minimum between $19^{th}$ to $23^{th}$ February. This highest and lowest sea ice area timing are around 40 days later when the temperatures reach minimum and maximum respectively\footnote{Source: mean temperature for 3 different stations of Antarctic Ocean cane be found here: \url{http://www.coolantarctica.com/Antarctica\%20fact\%20file/antarctica\%20environment/vostok_south_pole_mcmurdo.php}}. So the time lag between extreme temperature and extreme sea ice area of Arctic is around 40 days. The average sea ice area of Arctic and Antarctic Ocean is 11.50, 11.76 million square Km respectively. Now in the following section we present the analysis of mean and variance function.

\subsection{Mean Function Analysis}
\begin{figure}[h!]
\centering
\caption{Mean curves (decade wise) of Arctic and Antarctic sea ice area (in million square km) after smoothing. Red curve represents the mean function for the 1979-1991, Green represnts for 1992 - 2003 and Blue represents the mean function for the 2004 - 2015}
\label{fig5}
\includegraphics[height=3.5in]{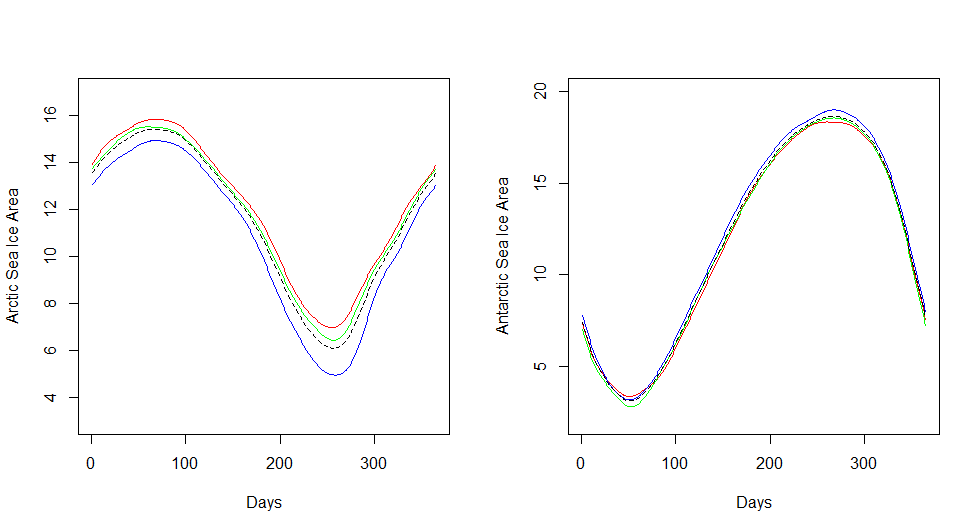}
\end{figure}

In FDA, the mean function is defined as
$$\disp\overline{x}(t) = \frac{1}{N}\sum_{i=1}^{N}x_{i}(t), $$ where $\overline{x}(t)$ is the pointwise mean of the given $N$ functions. We present the mean function of sea ice area for both Arctic and Antarctic oceans in figure \ref{fig5}. The dashed line in figure \ref{fig5} represents the mean function of smoothed sea ice area computed over all 37 years.The red curve represents mean function for both Arctic and Antarctic sea ice area for 13 years (1979-1991). Green and Blue line represents mean function for both Arctic and Antarctic sea ice area for (1992-2003) and (2004-2015) respectively.

For the Arctic Ocean, the minimum sea ice area (which occurred in between $50^{th}$ - $54^{th}$ day of the year) is respectively 6.96, 6.43, 4.95 million square km for the three consecutive decades respectively. The maximum sea ice area (which occurred in between $265^{th}$ - $269^{th}$  day of the year) is respectively 15.83, 15.51, 14.92 million square Km for the three consecutive decades respectively. For Antarctic Ocean, the minimum sea ice area (which occurred in between $253^{th}$ - $257^{th}$ day of the year) is respectively 3.34, 2.78, 3.16 million square Km for the three consecutive decades respectively. The maximum sea ice area (which occurred in between $64^{th}$ - $68^{th}$  day of the year) is 18.35, 18.57, 19.01 million square Km for the three consecutive decades respectively. Clearly, from the decade wise mean function curve, there is a downward trend in Arctic sea ice area. However, unless we look into $95\%$ confidence interval of the curve, we cannot conclude that this trend is statistically significant. However, for the Antarctic ocean, visually there is no trend.

\begin{figure}
\centering
\caption{Consecutive difference of deacde wiae mean curves for both Arctic and Atlanctic Oceans.}
\label{fig6}
\includegraphics[height=3in]{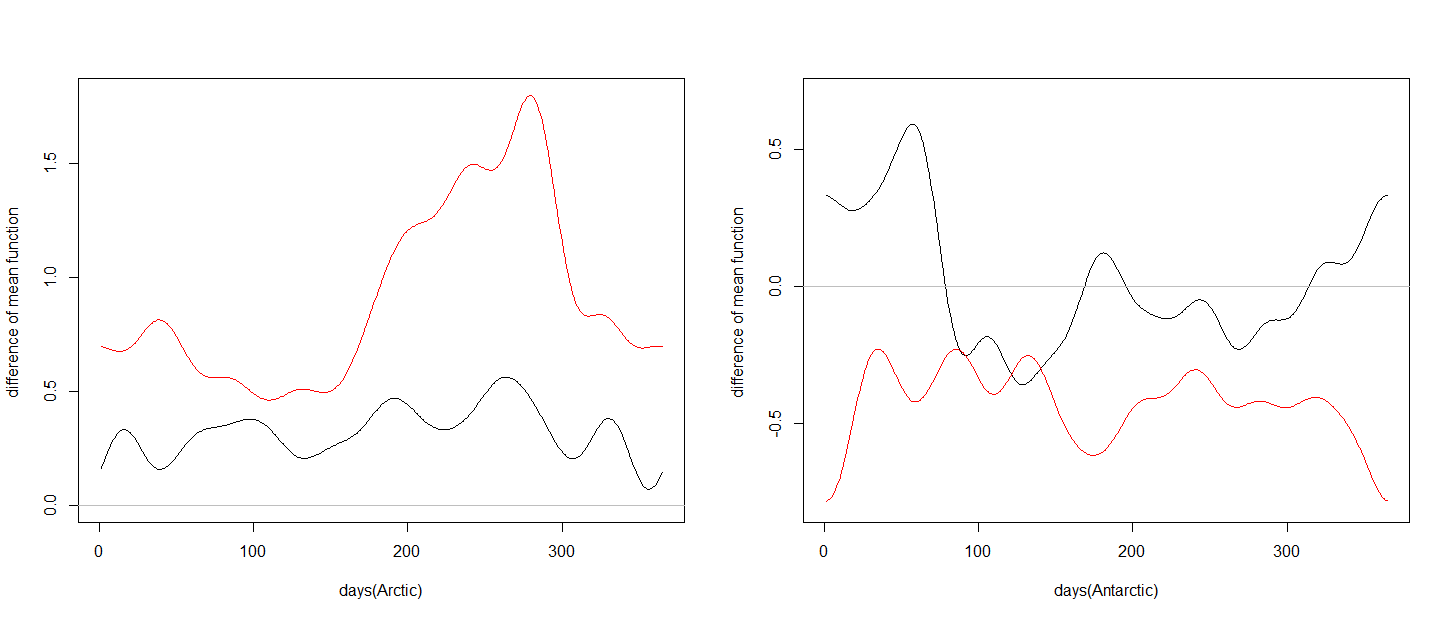}
\end{figure}

In figure \ref{fig6}, we present the difference of mean functions between first two decades (i.e., the difference between green and red curve in figure \ref{fig5}) by the black curve. Similarly, we present the difference of mean functions between last two decades (i.e., difference between blue and green curve in figure \ref{fig5}) by red curve. In the Arctic Ocean both the curve are in the positive region but in the Antarctic ocean both the difference curves are hovering around zero line. During the $170^{th}$ to $300^{th}$ day of the year  the change in sea ice area (in the Arctic ocean) is more prominent than the rest of the year.

\subsection{Variance Function Analysis}
\begin{figure}[h!]
\centering
\caption{Variance curves (decade wise) of Arctic and Antarctic sea ice area (in million square km) after smoothing. Red curve represents the mean function for the 1979-1991, Green represnts for 1992 - 2003 and Blue represents the mean function for the 2004 - 2015}
\label{fig7}
\includegraphics[ height=3in]{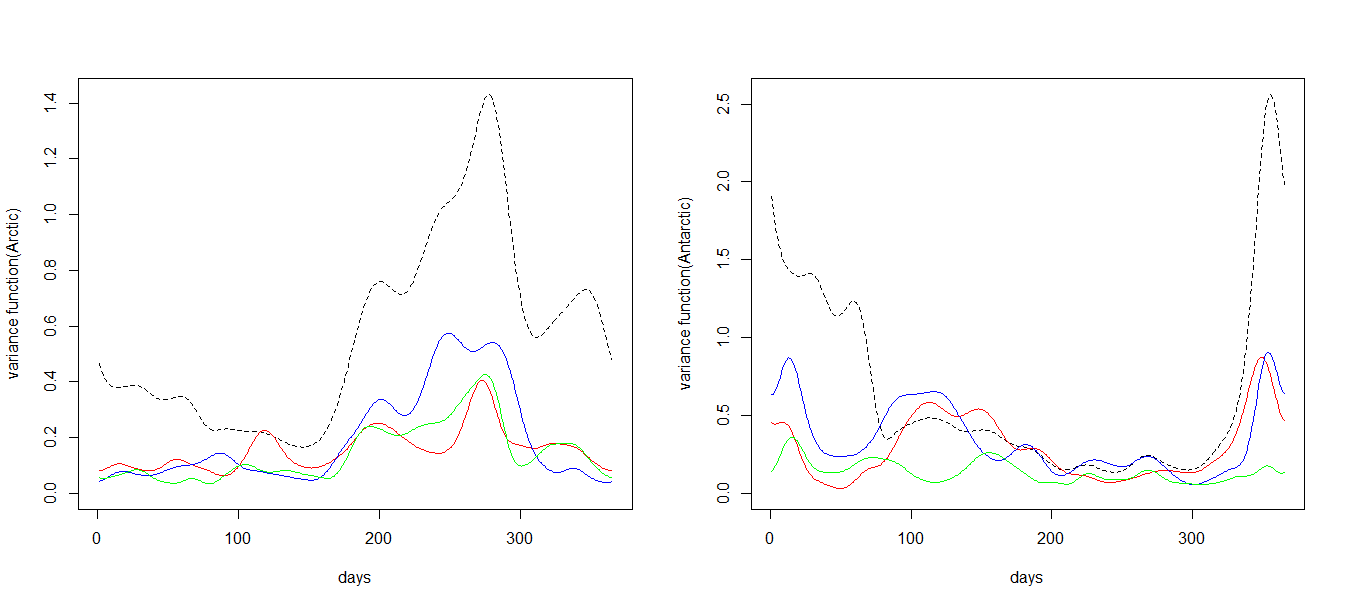}
\end{figure}

In FDA, the variance function (\cite{RamsaySilverman2005}) is defined as follows
$$
Var(X_{t}) = \dfrac{1}{N-1} \sum_{i=1}^{N}[x_{i}(t)-\overline{x}(t)]^2 ,
$$
where $Var(X_{t})$ is the pointwise variance of the given $N$ functions (for our case $N$ = number of years considered). In figure \ref{fig7}, we present the variance functions corresponding to the year slots as in mean function slot for both Arctic and Antarctic Oceans. For the Arctic Ocean in the day interval $[170,300]$ the variance is much higher for all the curves. However, for rest of the year, the variance is very low. So We have a non-constant variance indicating high volatility during summer. For Antarctic Ocean, the variance is very low throughout the year and also has no significant time effect during the year.

\section{Sea Ice Area Analysis via Bootstrap\label{sec5}}
We divide 37 years into $t$ many blocks named as $d_{1},d_{2},\cdots,d_{t}$, where we took $t$ to be 2, 3, 4, 5 respectively. Now for given $t$, the $i^{th}$ block $d_{i}$ where $i \in \{1,2,\cdots,t\}$,  we compute the $95\%$ bootstrap  confidence band (\cite{bootstrap1994}) of the block $d_{i}$ along with its bootstrap mean function.

\subsection{Boostrap Mean Function and Confidence Band Computation} 

Suppose block $d_{i}$ consists of $n_i$ many years. Then for the block $d_{i}$, we draw $n_{i}$ many resamples with replacement. We compute the mean function of the resampled curves. We repeat this resampling process for $B$ many time, i.e., we consider $B$ to be the bootstrap sample size and compute $B$ many bootstraps mean functions. Then we took pointwise mean, variance and $95\%$ confidence band of these $B$ mean functions. This is the estimated bootstrap mean function of the decade $d_{i}$ and the pointwise $95\%$ bootstrap confidence band. Throughout the paper, we consider $B$ to be $5000$.

\subsection{Analysis for Arctic ocean}

\begin{figure}[h!]
\centering
\caption{Red Line: Year 1979-1996.
Blue Line: Year 1997-2015.}
\label{fig8}
\includegraphics[height=3in]{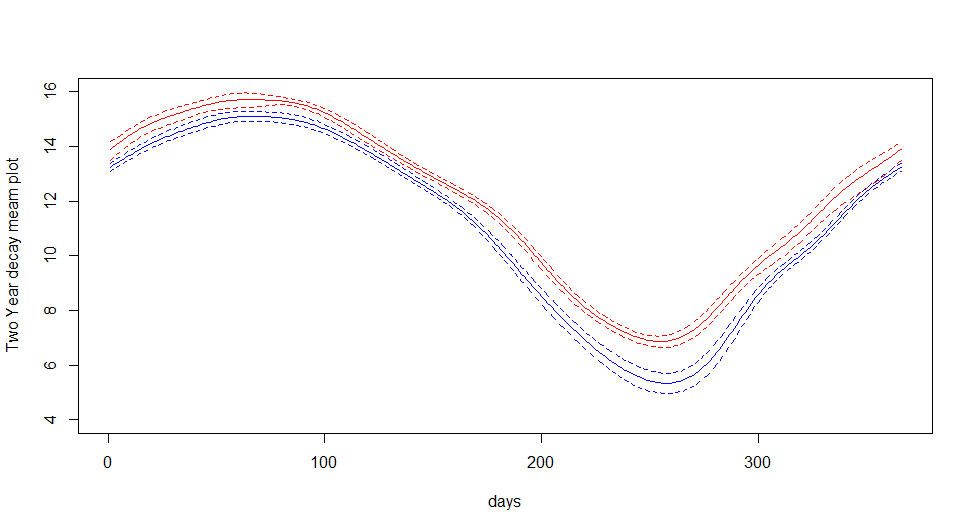}\\
\caption{Red Line: Year 1979-1990.
  Green Line: Year 1991-2001.
  Blue Line: Year 2002-2015.}
\label{fig9}
\includegraphics[height=3in]{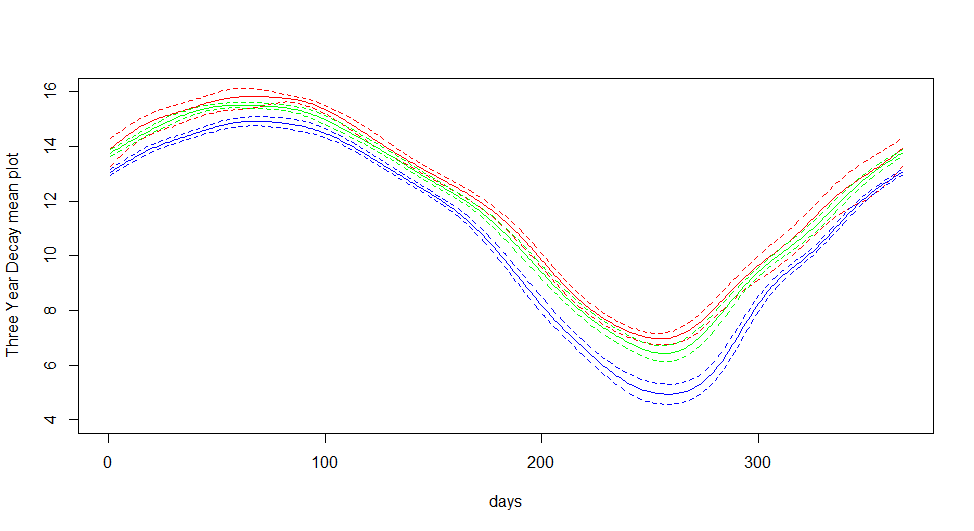}\\
\caption{Red Line: Year 1979-1987.
  Green Line: Year 1988-1996.
  Brown Line: Year 1997-2005.
  Blue Line: Year 2006-2015.}
\label{fig10}
\includegraphics[height=3in]{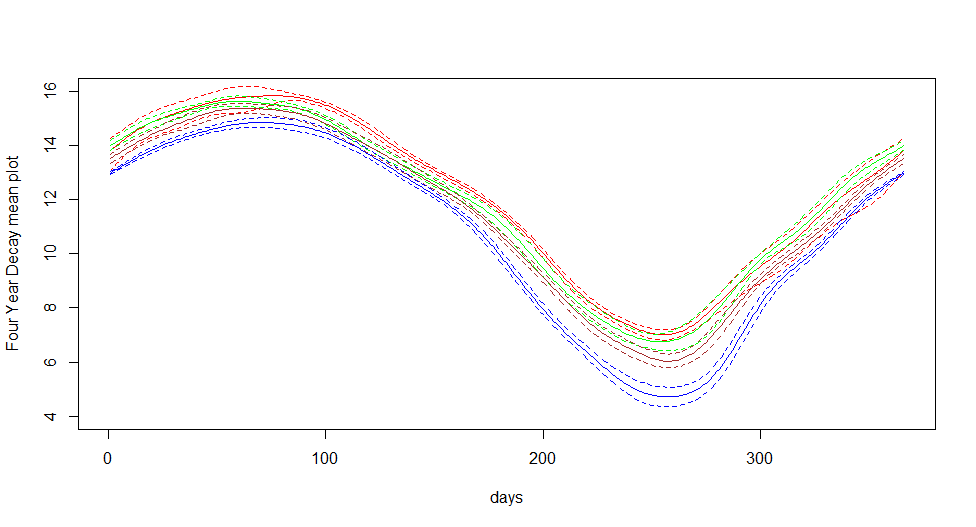}\\
\end{figure}

\begin{figure}[h!]
\centering
\caption{Red Line: Year 1979-1985.
  Green Line: Year 1986-1992.
  Brown Line: Year 1993-1999.
  Yellow Line: Year 2000-2007.
  Blue Line: Year 2008-2015.}
\label{fig11}
\includegraphics[height=3in]{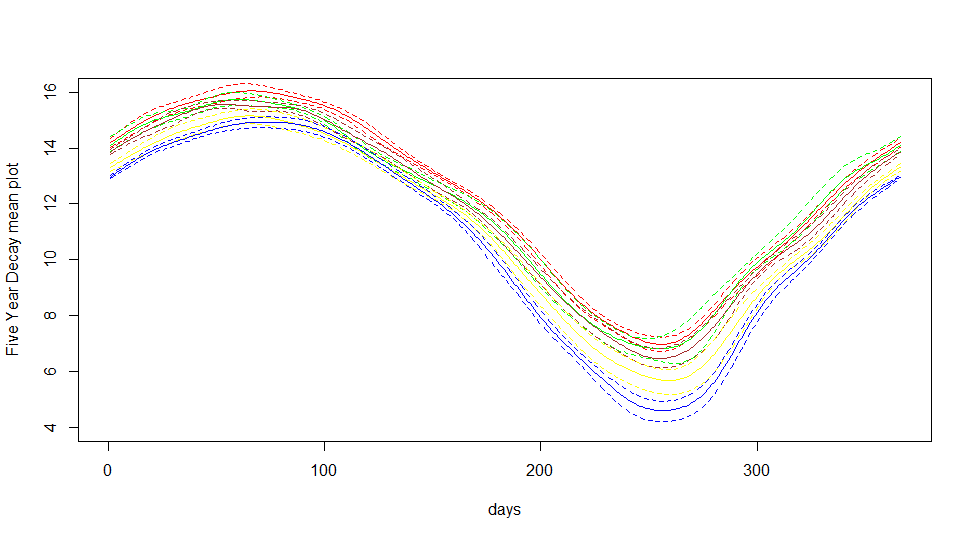}\\
\end{figure}

Here we present the confidence band for sea ice area of Arctic Ocean over different decades. For all colors, the dotted line shows the confidence band. The figure \ref{fig8} we present the analysis when $t=2$. That is we split the 37 years of data into two groups.The first group represents the first eighteen years of data from 1979 to 1996 and the second group represents the next nineteen years of data, i.e., from 1997 to 2015. It is visually clear that $95\%$ bootstrap confidence band for the two groups are non-overlapping. Hence we can conclude that mean function of recent group (1997-2015) is statistically significantly below the compare to mean curve estimated for 1979-1996. In figure \ref{fig9} we present the analysis when $t=3$. In this case, we split the 37 years of data into three groups. The first group represents the first twelve years of data from 1979 to 1990 (marked with red color). The second group consists of eleven years of data from 1991 to 2001 (marked with green color), and the third group consists of twelve years of data from 2002 to 2015 (marked with blue color). It is visually apparent that $95\%$ bootstrap confidence band for the three groups are non-overlapping. The drop in the confidence band is statistically significant.  In figure \ref{fig10} we present the analysis for $t=4$. In this case, we split the 37 years of data into four groups. The first group represents the first nine years of data from 1979 to 1987 (marked with red color). The second group consists of second nine years of data from 1988 to 1997 (marked with green color), the third group consists of third nine years of data from 1997 to 2005 (marked with brown color) and the fourth group represents the last ten years from 2006 to 2015. It is visually clear that $95\%$ bootstrap confidence band for the four groups are non-overlapping. The drop in the confidence band is statistically significant. In figure \ref{fig11} we present the analysis for $t=5$. In this case, we split the 37 years of data into five groups. The group represents (i) 1979--1985, (ii) 1986--1992, (iii) 1993--1999, (iv) 2000--2007 and (iv) 2008--2015. Each group is marked with red, green, brown, yellow and blue color respectively. Clearly, $95\%$ bootstrap confidence band for the recent years are non-overlapping with previous years. Now from figure \ref{fig8},\ref{fig9},\ref{fig10},\ref{fig11}, we can conclude that the Arctic Ocean has a downward trend and it is significant for all groups. If we focus on the three-decade plot i.e., figure \ref{fig9}, we observed that the mean curves are decreasing with time and are non-overlapping. In addition, we can see that the 95\% confidence band is also non overlapping throughout the year for the third group. So we have a statistically significant evidence of sea ice melting for Arctic Ocean. 

\subsection{Analysis for Antarctic Ocean}

\begin{figure}[h!]
\centering
\caption{Red Line:: Year 1979-1996.
Blue Line:: Year 1997-2015.}
\label{fig12}
\includegraphics[height=3in]{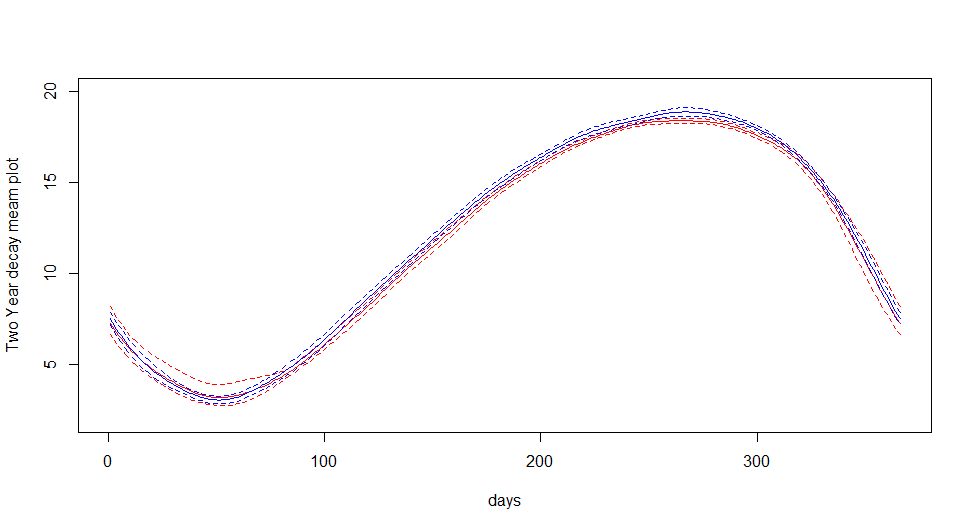}\\
\caption{Red Line:: Year 1979-1990.
  Green Line:: Year 1991-2001.
  Blue Line:: Year 2002-2015.}
\label{fig13}
\includegraphics[height=3in]{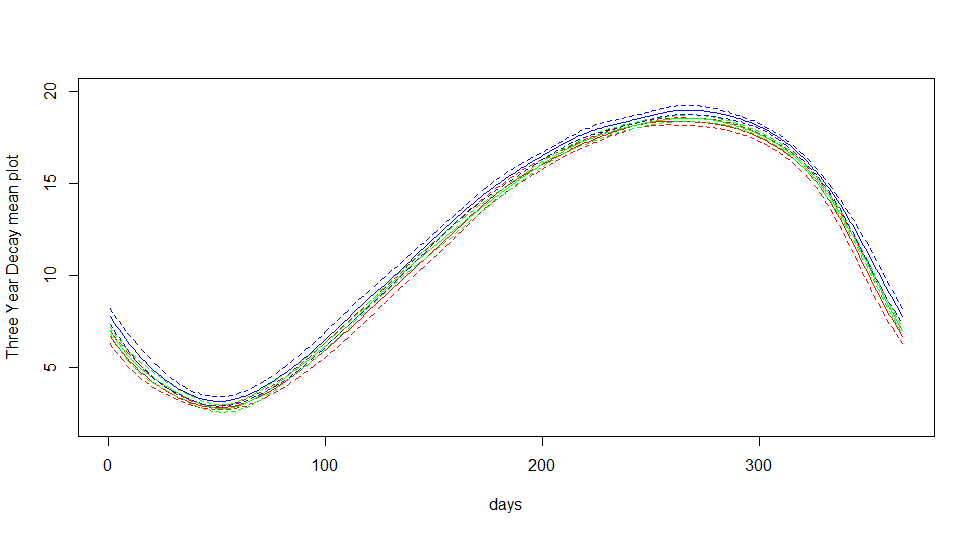}\\
\caption{Red Line:: Year 1979-1987.
  Green Line:: Year 1988-1996.
  Brown Line:: Year 1997-2005.
  Blue Line:: Year 2006-2015.}
\label{fig14}
\includegraphics[height=3in]{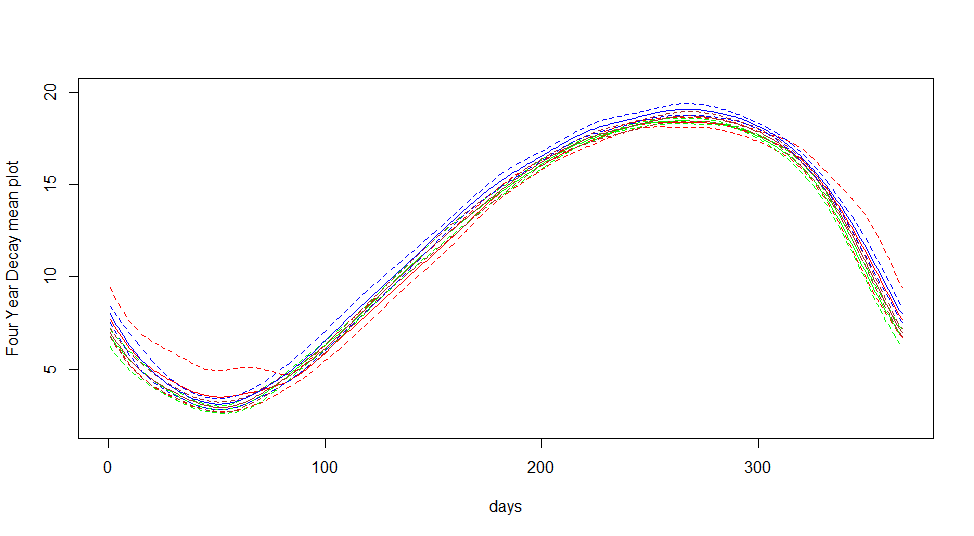}\\
\end{figure}

\begin{figure}[h!]
\centering
\caption{Red Line:: Year 1979-1985.
  Green Line:: Year 1986-1992.
  Brown Line:: Year 1993-1999.
  Yellow Line:: Year 2000-2007.
  Blue Line:: Year 2008-2015.}
\label{fig15}
\includegraphics[height=3in]{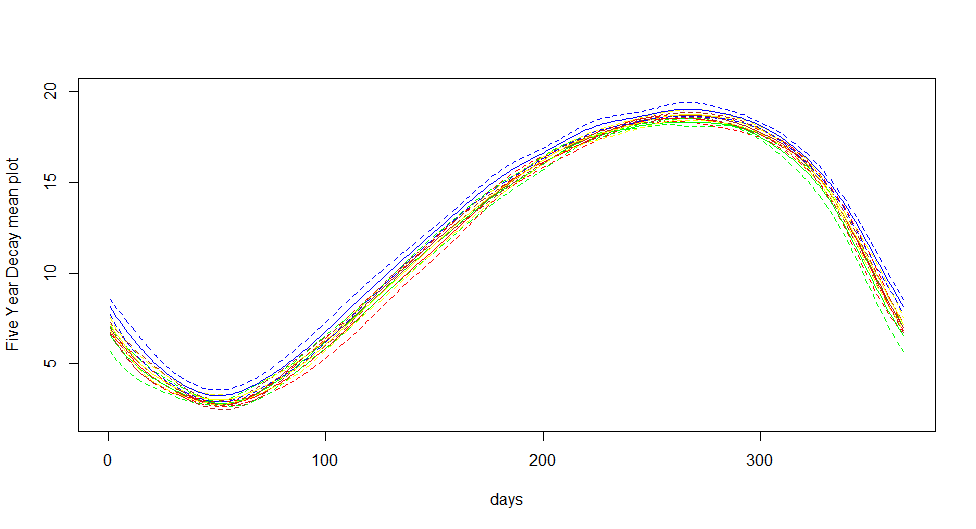}\\
\end{figure}

We consider the same strategy for the Antarctic Ocean as we present in the previous subsection for the Arctic Ocean. We present the same analysis for the Antarctic ocean for the figure \ref{fig12},\ref{fig13},\ref{fig14} and \ref{fig15}. In all the figures the $95\%$ bootstrap confidence bands for all the group are overlapping with each other. Hence we conclude that there is no statistical evidence of change in sea ice area in the Antarctic ocean.

\section{Phase Plane Analysis\label{sec6}}

In the Phase Plane analysis, we compare the sea ice area, its first and second difference operator for both Arctic and Antarctic Oceans. The first difference operator ($D$) on the smoothed Arctic (or Antarctic) sea ice area data is a good approximation of the derivative of the sea ice area. Also, it will be smooth and continuous as in the first level itself we have ignored the noise. So the second derivative of Arctic (or Antarctic) sea ice area can also be approximated by applying the difference operator ($D^{2}$) on the sea ice area. Now for our convenience, we will call the Sea ice area as Area, the first and second derivative as Velocity and Acceleration of this functional data. Note that all the quantities are  continuous and smooth.

\subsection{Area Vs. Velocity}

\begin{figure}[h!]
\centering
\caption{Area Vs. Velocity plot of Sea Ice Area for Arctic Ocean}
\label{fig16a}
\includegraphics[height=3in]{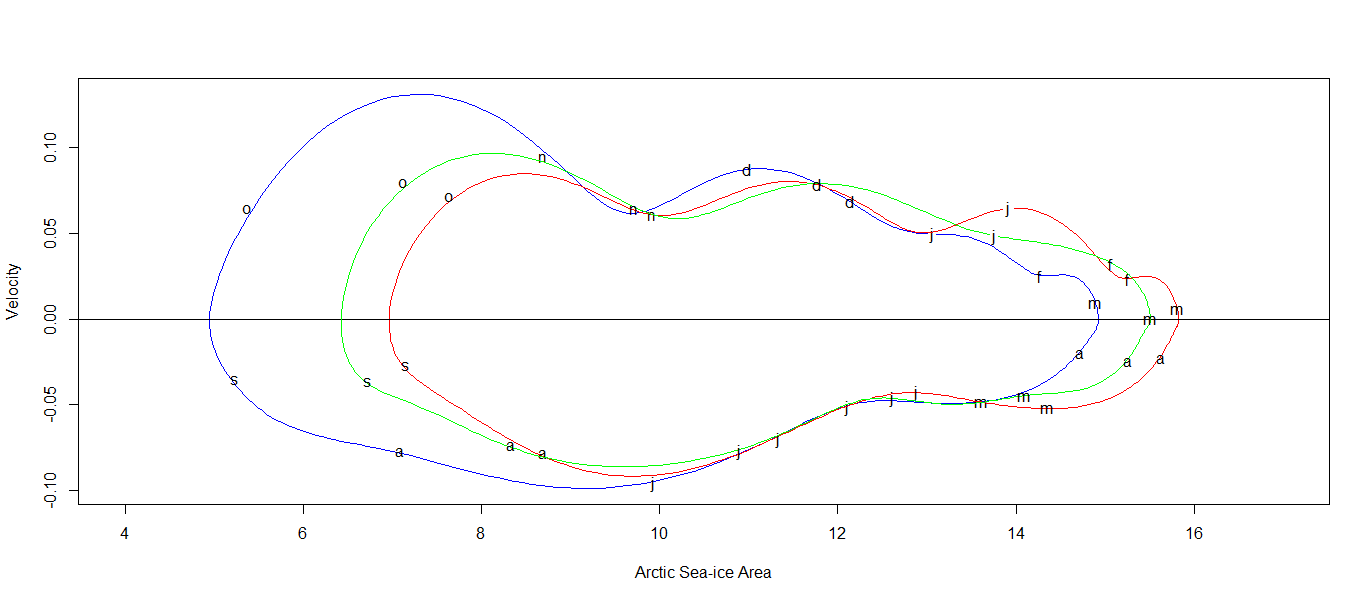}
\caption{Area Vs. Velocity plot of Sea Ice Area for Antarctic Ocean }
\label{fig16b}
\includegraphics[height=3in]{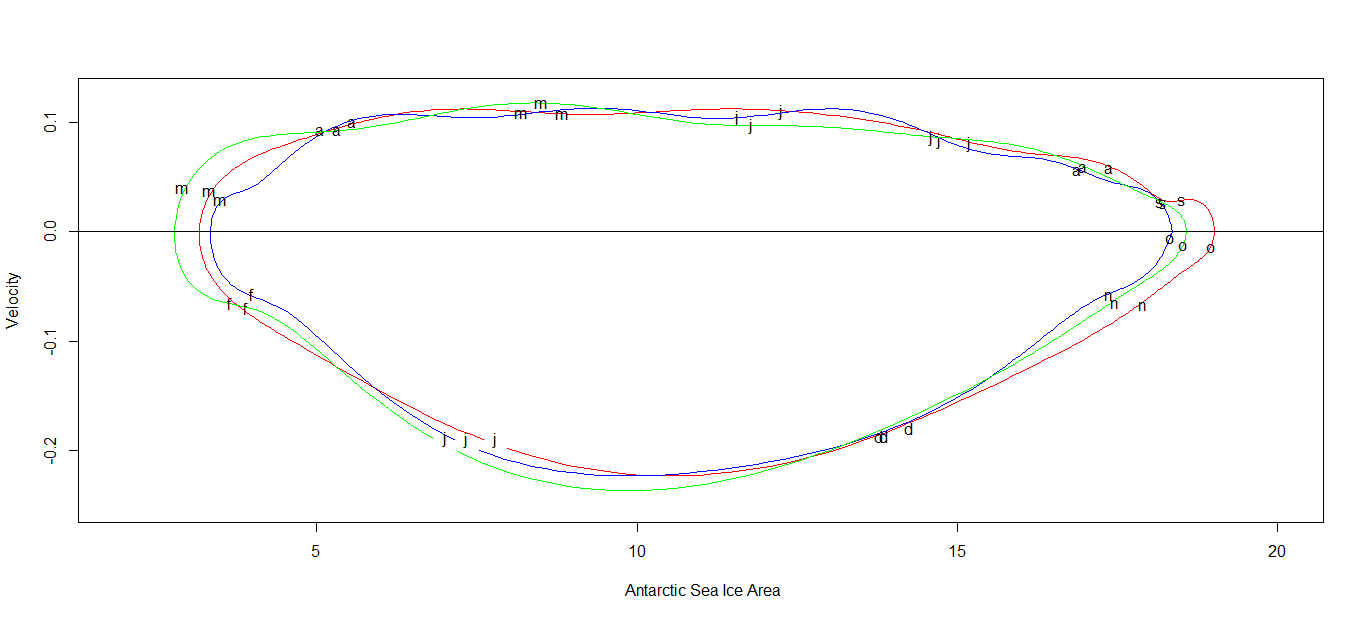}
\end{figure}

Here we consider the plot for the three-decade of Area Vs. Velocity for our analysis. The decades are defined as in the previous sections. One single line represents $365$ days phase plot. For our convenience, we tagged the first day of each month in each three-decade individually. We present the area versus velocity plot in figure \ref{fig16a} and \ref{fig16b}. Firstly, if we move from the first decade to the third one the whole curve has a left shift all over the year for the Arctic Ocean and when the sea ice area is maximum and minimum, the change in velocity is most significant. But in Antarctic Ocean (figure \ref{fig16b}), there is a slight change of phase plane for three decades is there but there is no right word or left word move of the phase plane. So the boundary of the phase planes is crossing each other many times. We can say we cannot see any significant change in the system.

Secondly, we observe that in between $1^{st}-15^{th}$ March the sea ice area reaches the maximum. And in between $5^{th}-15^{th}$ October the sea ice area reaches the minimum for all the three decades in the Arctic Ocean (figure \ref{fig16a}). Thirdly, in the time spell  May, June, November, December the velocity variation is mostly same throughout the decade wise observation for the Arctic Ocean. Fourthly, in the interval January to April the first decade graph is totally dominating the second, and second, is totally dominating the third one for the Arctic Ocean. The sea ice formation (and as a result melting, as nature always tries not to change its equilibrium) speed is going down in the period over decades when the sea ice area is maximum. 

Fifthly, during the time interval of July to October; when the sea ice area reaches the minimum, the second decade is dominating the first decade and dominated by the third one, i.e. sea ice area decreases more rapidly with the decades. Finally, The left shift for phase plane plot in the Arctic Ocean in the first and second decade is significantly less than that of second and third one. And also in the September the shift towards left is more than that of March. All the observations are Statistical evidence of a decrease in sea ice area.

\subsection{Analysis of Area versus Acceleration}
\begin{figure}[h!]
\centering
\caption{Area Vs. Acceleration plot of Sea Ice Area for Arctic Ocean}
\label{fig18}
\includegraphics[height=3in]{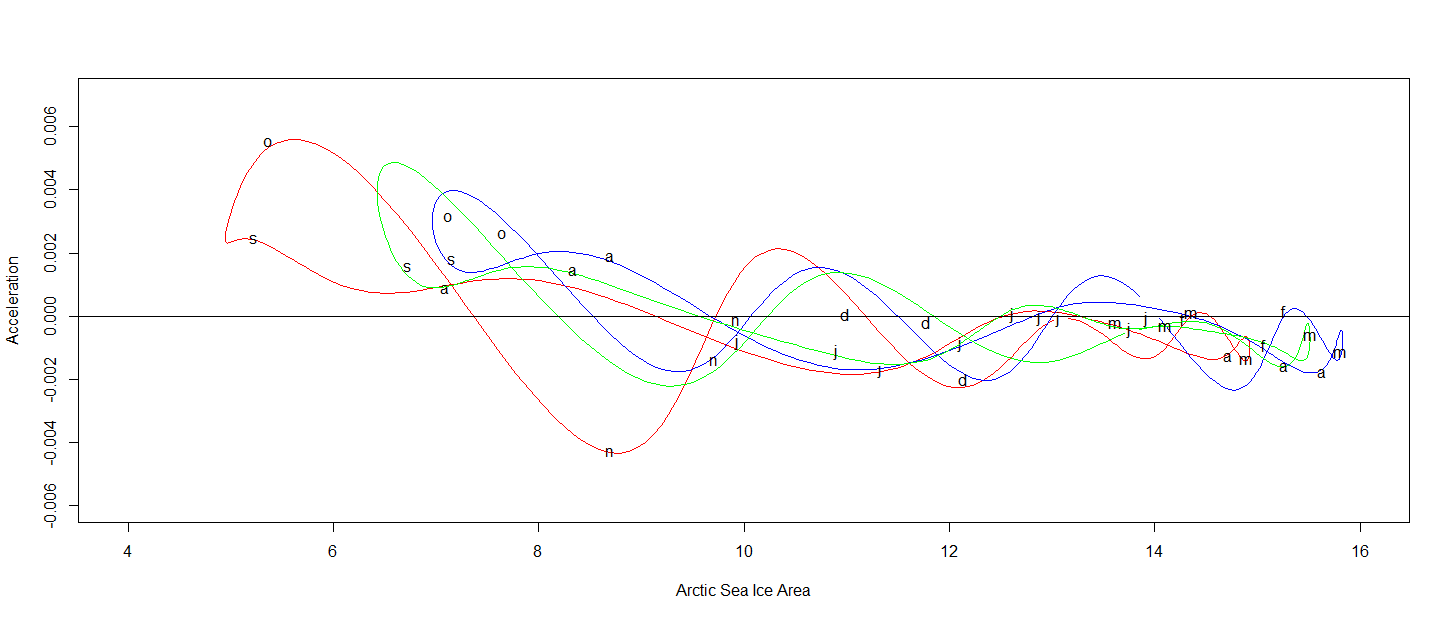}
\caption{Area Vs. Acceleration plot of Sea Ice Area for Antarctic Ocean}
\label{fig19}
\includegraphics[height=3in]{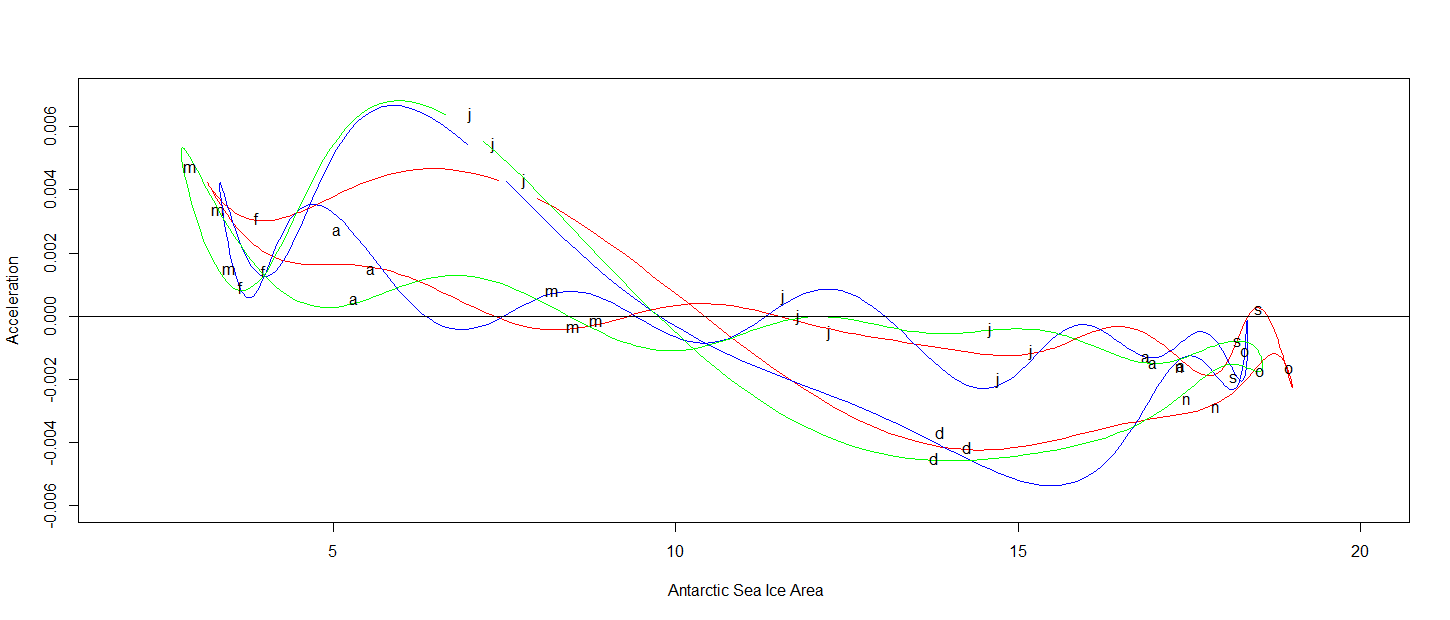}
\end{figure}

From the figures \ref{fig18} and \ref{fig19} below we can see that the Area-Acceleration plot has mostly the same trend with a time lag of $6$ months between Arctic and Antarctic Oceans. Due to space constraint, we just present the analysis about the Arctic Ocean only. In the Area versus Acceleration plot, each color code represents exactly same as it represents in the last phase plane plot, except we replace the velocity of sea ice area with acceleration. We labeled the first day of each month only on the third year decade; the months are in same places in the other curves also. From the figures \ref{fig18}, we can conclude that there is a left shift of the closed curve as we move decades for the Arctic Ocean. In the time interval January to June, when the sea ice area reaches its maximum, the acceleration is negative, and there is a shift towards top left decades, i.e. sea ice area increases and decreases in lesser rate, which causes less sea ice area when we move decades. In the Area Vs. Velocity plot we had a very clear dominance in March and September, but from the Area versus the acceleration plot exactly similar evidence we can draw only in the September to October time span. It indicates most of the changes in the Arctic Ocean are explained by velocity or first difference operator.

\subsection{Analysis of Velocity versus Acceleration}

\begin{figure}[h!]
\centering
\caption{Velocity Vs. Acceleration plot of Sea Ice Area for Arctic Ocean}
\label{fig20}
\includegraphics[height=3in]{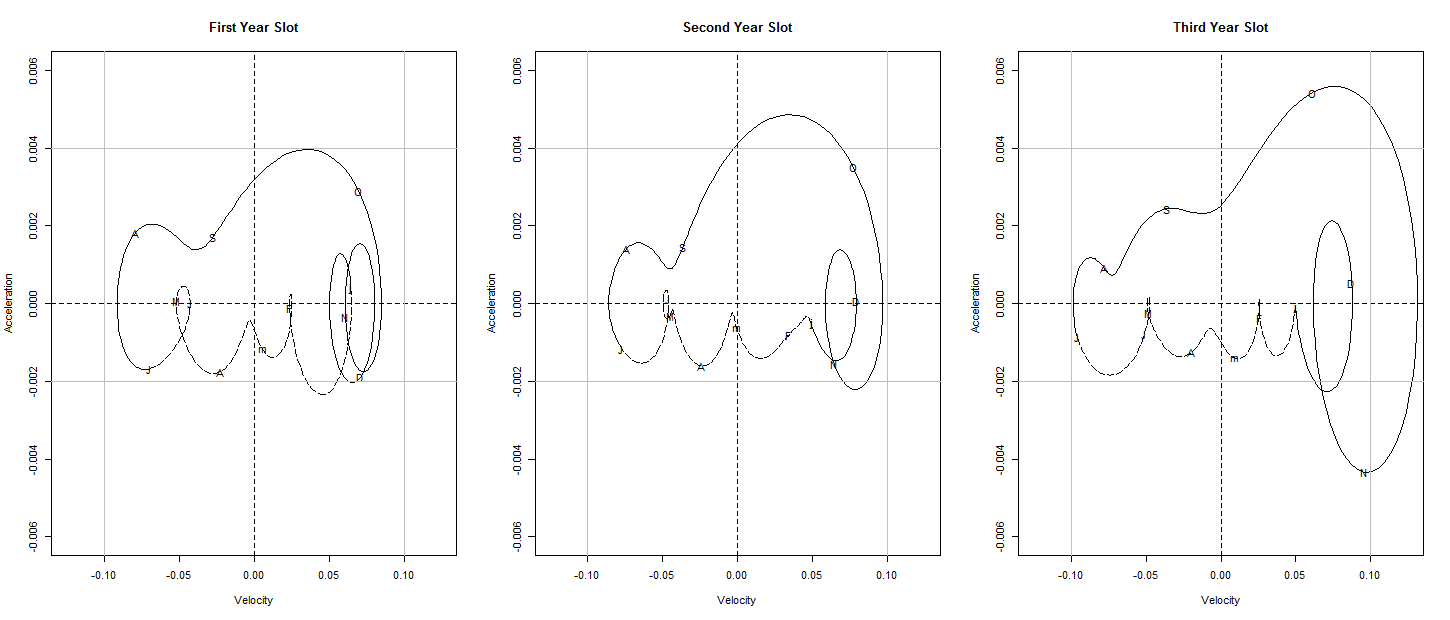}
\caption{Velocity Vs. Acceleration plot of Sea Ice Area for Antarctic Ocean}
\label{fig21}
\includegraphics[height=3in]{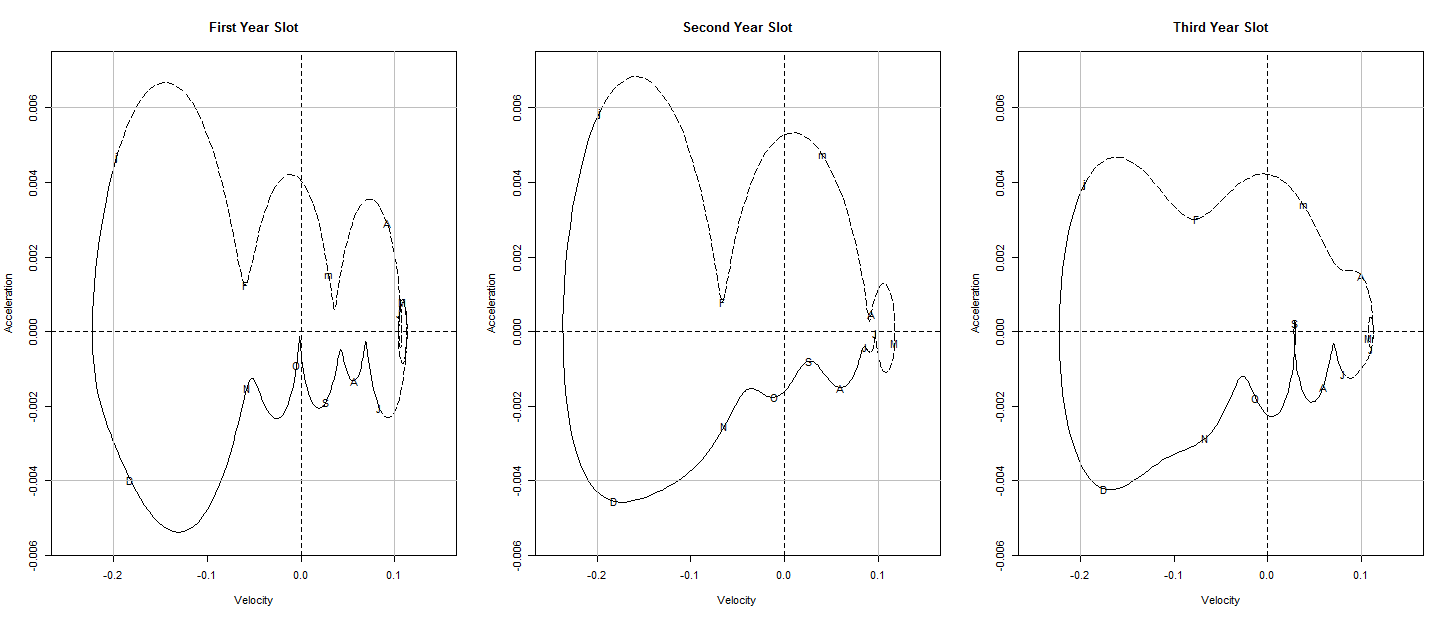}
\end{figure}

The velocity versus acceleration plot has a lot of physical interpretations. We can think of the velocity axis to be the potential energy and the acceleration axis to be the kinetic energy level of the system. From the velocity versus acceleration plot we can conclude as follows and for that purpose, we use the explanation from the reference(\cite{RamsaySilverman2005}, page 32).  The larger radius implies increasing energy transfer, which means the system is going away from equilibrium.  There is net positive velocity if it is right to the horizontal location of the center. Otherwise, it implies negative velocity. If it is above zero from the vertical location of the center, there is a net velocity increase; if below zero, there is a drop in velocity.  Variations in the shapes of the cycles from year to year. In figure \ref{fig20}, we present the velocity versus acceleration plot for the Arctic Ocean only. For Antarctic ocean, we have present the plots in figure \ref{fig21}, and the plots do not show any evidence for change sea ice area.

Symbol for the graphs : Letters indicate mid-months, with lowercase letters used for January and March. For clarity, the first half of the year is plotted as a dashed line and the second half as a solid line. Symbol reference is inbuilt \texttt{R} code of \texttt{FDA} package.
In the first-quadrant; the sea ice area increase at an increasing rate. In the second quadrant; the sea ice area decrease at an increasing rate. In the third quadrant; the sea ice area decrease at a decreasing rate. In the fourth quadrant; sea ice area increase at a decreasing rate.

In the October to December subcycle, the radius of the cycle is strictly increasing decadewise. Therefore more energy transfer in the event corresponds to decrease in sea ice area in this time span. Gradually horizontal span from origin to right side of the curve is increasing decadewise, i.e. net positive velocity is increasing. Hence the sea ice area decreasing more rapidly.  In May to August subcycle, the horizontal span of the left side of the plot is more-or-less same. So no remarkable change in this region.

\subsection{Measurement of Percentage Change\label{sec7}}
\begin{figure}[h!]
\centering
\caption{Percentage Change in Sea ice area of Arctic and Atlanctic Ocean}
\label{fig22}
\includegraphics[height=3.2in]{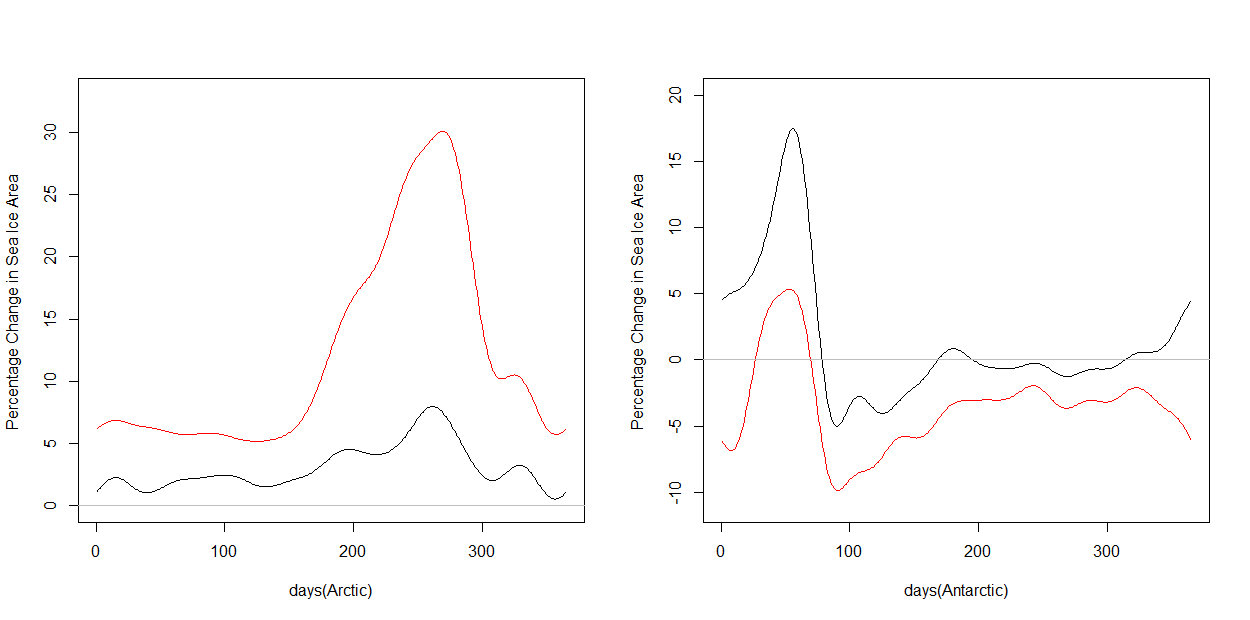}
\end{figure}

In the previous sections, we present the statistical evidence of decrease in sea ice area over decades in the Arctic Ocean. Here, we provide a measure of percentage change in  drop in sea ice area over the decade in the Arctic Ocean. Suppose $d_{1}$ ,$d_{2}$ ,$d_{3}$ be the mean sea ice area function of Arctic (Antarctic) Ocean for first, second and third decade. $d_{j}(i)$ represents the value evaluated at the point i of the continuous function $d_{j}$  for $i\in \{1,2,...365\}$ and $j\in \{1,2,3 \}$. We define two measures $\mu_{1,j}$ with $j\in \{2,3\}$ as follows:
\begin{equation*}
\mu_{1,j}(x)=\frac{(d_{j}-d_{1})}{d_{1}}(x)
\end{equation*}
The values of the $\mu_{1,j}$ evaluated at $i$ for $i \in {1,2,...365}$ is $\frac{(d_{j}-d_{1})}{d_{1}}$. This two measures $\mu_{1,2}$ and $\mu_{1,3}$ provide the percentage of change in the second and third decade with respective to the first one. We present the measurement of percentage change curve in figure \ref{fig22}. The red (black) curve presents the percentage change of sea ice area in both Arctic and Antarctic Oceans for the third (second) decade with respective to the first one. The figure indicates, during the summer, the current decades is observing the size of the sea ice area about 30\% less, than what it used to be during the first decade.

\section{Conclusions}
In this paper, we have provided how to model the sea ice area daily data for both Arctic and Antarctic region as functional data. Each year daily sea ice data has been considered as a smooth continuous function over time. We use the Fourier basis to estimate the smooth functions for $37$ years ($1989-2015$). In the paper, we also provide the optimal number of basis for the Fourier smooth. After analyzing the functional data, we got significantly strong Statistical evidence of decrease in sea ice area of Arctic Ocean whereas in Antarctic Ocean no Statistical evidence for sea ice cover melting has been found. For our analysis of sea ice area, we looked at the progressive blocks of years, the mean curve, the bootstrap confidence band ($95\%$) for each block of year for both Arctic and Antarctic Oceans. We find the statistical evidence of sea ice area melting of Arctic Ocean in each progressive block of years. We also find that the sea ice cover has decreased by $30\%$ during the summer time in the Arctic Ocean as compared to that of the sea ice cover change in late $80's$. All the similar analysis have been done for Antarctic Ocean also, but no significant statistical evidence of Sea ice area have been found over time for the observed $37$ years.      
%Also, we find $6$ seas out of $9$ of Arctic Ocean has significant Statistical evidence of decreasing sea ice area. $3$ seas out of $5$ seas of Antarctic Ocean has significant evidence of sea ice cover melting though as a whole for Antarctic Ocean no significant evidence is found. 

\bibliographystyle{plainnat}
\bibliography{Biblio-Database}

\newpage
\section*{Figures}

\newpage

\newpage

\newpage

\newpage

\end{document}